\RequirePackage{ifpdf}
\ifpdf % We are running pdfTeX in pdf mode
\documentclass[pdftex]{sigma}
\else
\documentclass{sigma}
\fi

\newcommand{\p}{\partial}
\newcommand{\vp}{\varphi}
\newcommand{\la}{\lambda}
\newcommand{\R}{\mathcal{R}}
\newcommand{\E}{\mathbb{E}}
\newcommand{\al}{\alpha}

\numberwithin{equation}{section}

\begin{document}

\allowdisplaybreaks

\renewcommand{\thefootnote}{$\star$}

\renewcommand{\PaperNumber}{095}

\FirstPageHeading

\ShortArticleName{On Darboux's Approach to $R$-Separability of Variables}

\ArticleName{On Darboux's Approach\\ to $\boldsymbol{R}$-Separability of Variables\footnote{This paper is a
contribution to the Special Issue ``Symmetry, Separation, Super-integrability and Special Functions~(S$^4$)''. The
full collection is available at
\href{http://www.emis.de/journals/SIGMA/S4.html}{http://www.emis.de/journals/SIGMA/S4.html}}}

\Author{Antoni SYM~$^\dag$ and Adam SZERESZEWSKI~$^\ddag$}

\AuthorNameForHeading{A.~Sym and A.~Szereszewski}

\Address{$^\dag$~Department of Mathematical Methods in Physics,
          Faculty of Physics, \\
          \hphantom{$^\ddag$}~University of Warsaw,  Poland}
\EmailD{\href{mailto:Antoni.Sym@fuw.edu.pl}{Antoni.Sym@fuw.edu.pl}}

\Address{$^\ddag$~Institute of Theoretical Physics, Faculty of Physics, University of Warsaw,  Poland}
\EmailD{\href{mailto:aszer@fuw.edu.pl}{aszer@fuw.edu.pl}}

\ArticleDates{Received February 18, 2011, in f\/inal form October 02, 2011;  Published online October 12, 2011}

\Abstract{We discuss the problem of $R$-separability (separability of variables with a factor $R$) in
the stationary Schr\"odinger equation on $n$-dimensional Riemann space. We follow the approach of Gaston Darboux
who was the f\/irst to give the f\/irst general treatment of $R$-separability in PDE (Laplace equation on~$\E^3$). According to Darboux $R$-separability amounts to two  conditions: metric is isothermic
(all its parametric surfaces are isothermic in the sense of both classical dif\/ferential geometry and
modern theory of solitons) and moreover when an isothermic metric is given their Lam\'e coef\/f\/icients
satisfy a single constraint which is either functional (when $R$ is harmonic) or dif\/ferential
(in the opposite case). These two conditions are generalized to $n$-dimensional case. In particular we def\/ine
$n$-dimensional isothermic metrics and distinguish an important subclass of isothermic metrics which we call
binary metrics. The approach is
illustrated by two standard examples and two less
standard examples.
In all cases the approach of\/fers alternative and much simplif\/ied proofs or derivations. We formulate
a systematic procedure to isolate $R$-separable metrics. This procedure
is implemented in the case of  3-dimensional Laplace equation. Finally
we discuss the class of Dupin-cyclidic metrics which are non-regularly $R$-separable in the Laplace
equation on~$\E^3$.}

\Keywords{separation of variables; elliptic equations; diagonal $n$-dimensional metrics; isothermic surfaces;
Dupin cyclides; Lam\'e equations}

\Classification{35J05; 35J10; 35J15; 35Q05; 35R01; 53A05}

\section{Introduction}\label{section1}

One of the highlights of Darboux's research on the whole is a memoir~\cite{D6} devoted mainly to
orthogonal coordinates in Euclidean spaces. The fundamental monograph~\cite{D8} includes much of the
material of~\cite{D6}. The last f\/ifty pages of the third and last part of the memoir~\cite{D7}
are nothing else but the f\/irst general treatment of  the $R$-separability of variables (separability of
variables with a factor~$R$) in a PDE.

\subsection[$R$-separability setting]{$\boldsymbol{R}$-separability setting}\label{section1.1}

Let
 \begin{gather}
                                \Lambda\psi=0    \label{1.1}
 \end{gather}
be a linear PDE in variables  $x=\left(x^1,x^2,\dots,x^n\right)$  for an unknown (function)
$\psi(x)$ and of order~$N$.

 \begin{definition}\label{definition1}
PDE~(\ref{1.1}) is $R$-separable (into ODEs) or $x$-variables are $R$-separable in equation~(\ref{1.1}) if
there exist a non-zero function $R(x)$ and $n$ linear ODEs
 \begin{gather}
                     L_i \vp_i  = 0 , \qquad   i  =  1,2,\dots,n ,	    \label{1.2}
 \end{gather}
each of order $\nu_i\leq N$ and  for a function  $\vp_i(x^i)$ such that  the following implication holds
 \begin{gather}
  \text{if}\quad  L_i\vp_i=0, \quad   i  =  1,2,\dots,n ,\qquad  \text{then}\quad \psi(x) = R(x)\prod_ i\vp_i\big(x^i\big)\quad
  \text{solves (\ref{1.1})}.    \label{def1}
 \end{gather}
 \end{definition}
Also we say that $R$-separation occurs in equation~(\ref{1.1}). Equations~(\ref{1.2}) are called separation equations.
\begin{remark}\label{remark1}
Following Darboux we assume that coef\/f\/icients of each equation~(\ref{1.2}) are just functions of variable~$x^i$
not necessarily dependent on extra variables (parameters). This freedom from the St\"ackel imperative is essential. Hence if~(\ref{def1}) holds then we have a family of
solutions to (\ref{1.1}) depending at least
on $\sum_i\nu_i$ parameters.
\end{remark}

\begin{remark}\label{remark2}
If  $R=1$ or  more generally if  $R =\prod_i r_i(x^i)$, we replace the term ``$R$-separability'' by
the term ``separability''.
\end{remark}

\subsection[$R$-separability in the Schr\"odinger equation]{$\boldsymbol{R}$-separability in the Schr\"odinger equation}\label{section1.2}

We assume that a Riemann space $\R^n$ admits local orthogonal coordinates $u=(u^1,\dots,u^n)$ in which
the metric has the following form
\begin{gather}
   ds^2=\sum_{i=1}^n H_i^2 \big(du^i\big)^2.   \label{metdiag}
 \end{gather}

H.P.~Robertson was the f\/irst to consider the stationary Schr\"odinger equation on $\mathcal{R}^n$
equipped with orthogonal coordinates
 \begin{gather}
    \Delta \psi+\big(k^2-V\big)\psi=0 ,  \label{Schr}
 \end{gather}
where
 \begin{gather*}
     \Delta= h^{-1}\sum_{i=1}^n\frac{\p}{\p u^i}\frac{h}{H_i^2}\frac{\p}{\p u^i},
   \qquad  h=H_1H_2\cdots H_n   %\label{Delta}
 \end{gather*}
is the Laplace--Beltrami operator on $\mathcal{R}^n$, $k$ is a scalar and $V=V(u)$ is a potential function~\cite{HR}.

We adapt the Def\/inition~\ref{definition1}  to the case of  equation (\ref{Schr}) as follows.
 \begin{definition}\label{definition2}
  The Schr\"odinger equation is $R$-separable or metric (\ref{metdiag})  and potential $V$ are $R$-separable
in the Schr\"odinger equation  if there exist
$2n+1$ functions $R(u)$ and $p_i(u^i)$, $q_i(u^i)$ $(i=1,2,\dots,n)$ such that the following
implication holds
 \begin{gather}\label{def2}
  %\bein{split}
  \vp''_i+p_i\vp'_i+q_i\vp_i=0,  \ \   i=1,2,\dots,n   \quad \Rightarrow \quad
   \psi(u)=R(u)\prod_i \vp_i\big(u^i\big) \ \
                    \textrm{solves\ \  (\ref{Schr})}.
 \end{gather}
  \end{definition}
Particular cases of equation (\ref{Schr}) are
\begin{itemize}\itemsep=0pt
  \item[a)] $n$-dimensional Laplace equation ($k=0$ and $V=0$)
     \begin{gather*}
     \Delta \psi=0 ,  %\label{Lapl}
    \end{gather*}
  \item[b)] $n$-dimensional Helmholtz equation ($V=0$)
    \begin{gather}
     \Delta \psi+k^2\psi=0 ,  \label{Helm}
    \end{gather}
  \item[c)] $n$-dimensional Schr\"odinger equation with $k=0$
    \begin{gather}
     \Delta \psi=V\psi .  \label{Schrk0}
    \end{gather}
 \end{itemize}

In the context of the $R$-separability  in the Schr\"odinger equation the following problem seems to be fundamental.

{\bf $\boldsymbol{R}$-separability problem.}
Let $\R^n$ be a Riemann space with a metric $ds^2=g_{ij}dx^i dx^j$, where $(x^i)$ are local coordinates.
We assume that $\R^n$ admits orthogonal coordinates and we are given  a~class of $R$-separable metrics~(\ref{metdiag}).
By $R$-separability problem, we mean the problem of isolating those metrics of the class which are equivalent to the
metric~$ds^2$.

 \begin{remark}\label{remark3}
As is well known a generic $\mathcal{R}^n$ for $n>3$ does not admit orthogonal coordinates \cite[p.~470]{LB}.
Any analytic $\mathcal{R}^3$ always admits orthogonal coordinates~\cite{EC} and even more any $\mathcal{R}^3$
of $C^\infty$-class also admits orthogonal coordinates~\cite{DT}. Also the problem when a given metric is
diagonalizable seems to be very dif\/f\/icult~\cite{KT}.
 \end{remark}

Robertson proved in \cite{HR} that any $n$-dimensional St\"ackel metrics satisfying the so called Robertson condition
is separable in the Schr\"odinger equation~(\ref{Schr}) (see also \eqref{kV} of this paper). The corresponding $R$-separability problem for $n$-dimensional
Euclidean space has been solved by L.P.~Eisenhart~\cite{LE2}.

\subsection[Darboux's  $R$-separability problem]{Darboux's  $\boldsymbol{R}$-separability problem}\label{section1.3}

Gaston Darboux was interested in $R$-separability of variables in the  Laplace equation on $\E^3$. His pioneering
research in the f\/ield of $R$-separability \cite{D4,D5,D7,D8} has been almost completely forgotten.
It can be interpreted as an advanced attempt to solve the following specif\/ic $R$-separability problem.

Here we do not use the original Darboux's notation dating back to Lam\'e writings. Instead, we apply the notation
used in this paper.

\begin{theorem}\label{theorem1}
The $3$-dimensional diagonal  metric
 \begin{gather}
   ds^2=H_1^2(u)\big(du^1\big)^2+H_2^2(u)\big(du^2\big)^2+H_3^2(u)\big(du^3\big)^2 \label{3dimdiag}
 \end{gather}
is $R$-separable in  $3$-dimensional Laplace equation
 \begin{gather}
   \left(\sum_{i=1}^3\partial_i\frac{H_1 H_2 H_3}{H_i^2}\p_i\right)\psi=0  \label{3dimLap}
 \end{gather}
if and only if the following two conditions  are satisfied
    \begin{gather}
  i) \quad   H_1=\frac{G_{(2)} G_{(3)}}{R^2f_1}, \qquad H_2=\frac{G_{(1)} G_{(3)}}{R^2f_2}, \qquad
     H_3=\frac{G_{(1)} G_{(2)}}{R^2f_3},           \label{H1H2H3}
    \end{gather}
where  $G_{(i)}$ does not depend on $u^i$ and  $f_i$ depends only on $u^i$,
     \begin{gather}
    ii)\quad     \sum_{i=1}^3 G_{(i)}^2f_i^2\left[\p^2_iR^{-1}+\frac{f'_i}{f_i}\p_i R^{-1}+q_i R^{-1} \right]=0
           \label{Reqn3dim}
    \end{gather}
for appropriately chosen functions  $q_i(u^i)$.

Moreover, the resulting separation equations  are
 \begin{gather*}
        \vp''_i+\frac{f'_i}{f_i}\vp'_i + q_i\vp_i=0, \qquad  i = 1,2,3 . %\label{3dimsepeqns}
 \end{gather*}
\end{theorem}

 \begin{remark}\label{remark4}
Theorem~\ref{theorem1} has never been explicitly stated by Darboux. Actually he applied this theorem in many places
of his research. E.g.~(3) in Chapter~IV of~\cite{D8} is a special case of (\ref{H1H2H3}) while~(69) in
Chapter~V of~\cite{D8} is a special case of~(\ref{Reqn3dim}).
 \end{remark}

In view of the Theorem~\ref{theorem1} the question of $R$-separability of variables in the Laplace equation
on~$\E^3$ amounts to the following  $R$-separability problem: to isolate all the metrics with Lam\'e coef\/f\/icients~(\ref{H1H2H3}) which are f\/lat and which satisfy~(\ref{Reqn3dim}). In other words, f\/irstly, one has to f\/ind
(classify) all solutions to the  Lam\'e equations ($i,j,k=(1,2,3), (2,3,1), (3,1,2)$)
 \begin{gather}
   H_{i,jk}-\frac{1}{H_j}H_{i,j}H_{j,k}-\frac{1}{H_k}H_{i,k}H_{k,j}=0 , \label{Lameoffdiag}\\
  \left(\frac{1}{H_i}H_{j,i}\right)_{,i}+\left(\frac{1}{H_j}H_{i,j}\right)_{,j}+\frac{1}{H_k^2} \label{Lamediag}
  H_{i,k}H_{j,k}=0 ,
 \end{gather}
under the ansatz (\ref{H1H2H3})  and, secondly, to select among them those satisfying the constraint~(\ref{Reqn3dim}).

Darboux was successful in solving the Lam\'e equations under the ansatz~(\ref{H1H2H3}). However as a rule he paid no
closer attention to the question of separation equations and thus with one exception the constraint~(\ref{Reqn3dim}) was not the subject of his detailed analysis. This exceptional case not covered by the modern
treatments of $R$-separability in the Laplace equation on~$\E^3$ is one of the Dupin-cyclidic metrics
\cite[pp.~283--286]{D8}. Indeed, Dupin-cyclidic metrics are non-regularly $R$-separable in the Laplace
equation on $\E^3$ and cannot be treated by the standard techniques discussed e.g.\ in~\cite{BKM}.
For a discussion of regular and non-regular $R$-separability see~\cite{WM2}.

 \begin{definition}\label{definition3}
A surface in $\E^3$ is isothermic if, away from umbilics, its curvature net can be conformally parametrized.
 \end{definition}

   Another important Darboux's result is as follows.

 \begin{theorem}\label{theorem2}
If the metric \eqref{3dimdiag} is $R$-separable in the Laplace equation on $\E^3$, then all the corresponding
parametric surfaces are isothermic.
 \end{theorem}

The class of isothermic surfaces is conformally invariant and  in particular includes
 \begin{itemize} \itemsep=0pt
  \item  planes and spheres,
  \item  surfaces of revolution,
  \item quadrics,
  \item  tori, cones, cylinders and their conformal images, i.e.\ Dupin cyclides,
  \item  cyclides or better Darboux--Moutard cyclides,
  \item  constant mean curvature surfaces and in particular minimal surfaces.
 \end{itemize}

Apart from the fact that the Theorem~\ref{theorem2} is a necessary condition for $R$-separability in the Laplace equation on $\E^3$,
it is an interesting connection between the linear mathematical physics (separation of variables) and the
non-linear mathematical physics (solitons). Indeed, the current interest in isothermic surfaces is mainly due
to the fact that their geometry is an important example of the so called integrable or soliton geometry~\cite{CGS,RS,FB,UHJ}.

 \begin{definition}\label{definition4}
 The metric (\ref{3dimdiag}) with Lam\'e coef\/f\/icients (\ref{H1H2H3}) is called isothermic.
 \end{definition}

\subsection{Aims and results of the paper}\label{section1.4}

In this paper we extend the original Darboux's approach to $R$-separability of variables in the Laplace equation on~$\E^3$ to the case of the stationary Schr\"odinger equation on $n$-dimensional Riemann space $\R^n$ admitting  orthogonal
coordinates.

The Darboux's Theorem~\ref{theorem1} is generalized as Theorem~\ref{theorem3}.  Correspondingly 3-dimensional isothermic metrics~(\ref{H1H2H3})
are  generalized to $n$-dimensional isothermic metrics~(\ref{2.16}) while 3-dimensional constraint (\ref{Reqn3dim})
is generalized to $n$-dimensional constraint~(\ref{eqn2}) which we call $R$-equation.

We distinguish a subclass (\ref{binmet}) of isothermic metrics which we call the binary metrics.
A~representative example of the binary metric is $n$-elliptic metric (\ref{ellmet}). In the case of a binary
metric the $R$-equation assumes the simpler form (\ref{th3}).

The approach is illustrated by examples of  the Section~\ref{section3}. Here we discuss two standard results  and two less
standard results. These are  1)~Robertson paper revisited  (Subsection~\ref{section3.1}), 2)~the $n$-elliptic metric  (Subsection~\ref{section3.2}) (standard results) and
3)~remarkable example of Kalnins--Miller revisited (Subsection~\ref{section3.3}), 4) f\/ixed energy $R$-separation revisited  (Subsection~\ref{section3.4}) (less standard
examples). In all cases the approach of\/fers alternative and simplif\/ied proofs or derivations.
%In particular a~complicated formalism developed in~\cite{CR} to treat the f\/ixed energy $R$-separation is now replaced by a single
%statement (see Proposition~\ref{proposition1} in Subsection~\ref{section3.4}).

The main result of the paper encoded in Theorem~\ref{section3} suggests the following procedure to identify a given
$n$-dimensional diagonal metric~(\ref{metdiag}) as $R$-separable in $n$-dimensional Schr\"odinger equation. The
procedure in question consists of three steps.

Firstly, we have to prove or disprove that the metric is isothermic. If the metric is not
isothermic, then it is not $R$-separable. Suppose it is isothermic. As a result this
step predicts $R$-factor and~$p_i$ coef\/f\/icients in the separation equations.
Secondly, we set up the correspon\-ding $R$-equation which we treat as an equation for~$q_i$ coef\/f\/icients in the separation equations.
Thirdly, we attempt to solve the $R$-equation. Any solution to the $R$-equation concludes the procedure:
$R$-separability of the starting metric is proved and in particular the corresponding separation equations
are explicitly constructed. Notice that unknowns $q_i$ enter into $R$-equation linearly and this is the right
place to introduce (linearly) extra parameters (separation constants) into the separation  equations.
In Subsection~\ref{section2.3} we introduce remarkable algebraic identities (B\^ocher--Ushveridze identities) which can
be successfully applied in solving $R$-equation.
This is a remarkably simple procedure and its implementation in the case of  3-dimensional Laplace equation is discussed
in Subsection~\ref{section4.1} together with the relevant examples.

Gaston Darboux found a class of Dupin-cyclidic metrics which are $R$-separable in the Laplace equation on~$\E^3$.
These are non-regularly $R$-separable and can not be covered by the modern standard approaches. In Subsection~\ref{section4.3} we re-derive this remarkable result. The original Darboux's calculations are long and rather dif\/f\/icult
to control. Here we simplify the derivation using the standard Riemannian tools (Ricci tensor and
Cotton--York criterion of conformal f\/latness).

\section[Isothermic metrics and $R$-equation]{Isothermic metrics and $\boldsymbol{R}$-equation}\label{section2}

\subsection{The main result}\label{section2.1}

Here we extend Darboux's Theorem~\ref{theorem1} valid for 3-dimensional Laplace equation (\ref{3dimLap}) to the case of $n$-dimensional
stationary Schr\"odinger equation (\ref{Schr}). Correspondingly, we extend the Def\/inition~\ref{definition4} of isothermic
metrics to $n$-dimensional case.

 \begin{theorem}\label{theorem3}
{\bf A.} The metric \eqref{metdiag} and the potential $V$ are $R$-separable in the Schr\"odinger equation~\eqref{Schr}
if and
only if the following two conditions are satisfied
 \begin{itemize}\itemsep=0pt
   \item first condition of $R$-separability
    \begin{gather}
     \left[\ln \left(R^2\frac{h}{H_i^2}\right)\right]_{,ij}=0, \qquad i\neq j, \label{th1_1}
    \end{gather}
   \item second  condition of $R$-separability  $($called  $R$-equation$)$
    \begin{gather}
     \Delta R + \left(k^2-V -\sum_{i=1}^n\frac{1}{H_i^2}q_i\right)R=0.  \label{th1_2}
    \end{gather}
 \end{itemize}

{\bf B.} The metric \eqref{metdiag} satisfies the first condition of $R$-separability if and only if
it can be cast
into the form
  \begin{gather}
    ds^2=R^{4/(2-n)}G_{(1)}^{2/(n-2)}G_{(2)}^{2/(n-2)}\cdots
        G_{(n)}^{2/(n-2)}\sum_{i=1}^n G_{(i)}^{-2}\frac{1}{f_i^2}\big(du^i\big)^2, \label{2.16}
  \end{gather}
where  $G_{(i)}$  does not depend on  $u^i$  while  $f_i$ depends only on $u^i$.

{\bf C.} If conditions \eqref{2.16} and \eqref{th1_2} are satisfied then the corresponding separation equations
read
 \begin{gather}
  \vp''_i+\frac{f'_i}{f_i}\vp'_i + q_i\vp_i=0.  \label{sepeqns}
 \end{gather}
 \end{theorem}

 \begin{proof}
{\bf A.} $R$-separability implies (\ref{th1_1}) and (\ref{th1_2}).

Indeed,  we insert
$\psi=R\prod_i\vp_i$ into (\ref{Schr}) and make use of (\ref{def2}). This results in
 \begin{gather}
   \sum_i\frac{1}{H_i^2}\left[\left(\ln R^2\frac{h}{H_i^2}\right)_{\!,i}-p_i\right]
    \frac{\vp'_i}{\vp_i}+R^{-1}\Delta R-\sum_i\frac{1}{H_i^2}q_i+k^2-V=0  \label{2.11}
 \end{gather}
for an arbitrary choice of solutions $\vp_i$. Let $(\vp_{i1},\vp_{i2})$
be a basis in the solution space of the corresponding equation. We put
 \begin{gather*}
   \vp_i=\la_i\vp_{i1}+\mu_i\vp_{i2}, \qquad \la_i,\mu_i={\rm const}.
 \end{gather*}
Thus for each $\vp_i$ $(\la_i\neq 0)$ we have
 \begin{gather}
   \frac{\vp'_i}{\vp_i}=\frac{\vp'_{i1}+\al_i\vp'_{i2}}{\vp_{i1}+\al_i\vp_{i2}},\label{2.12}
 \end{gather}
where $\al_i=\mu_i/\la_i={\rm const}$. Since (\ref{2.11}) with  $\frac{\vp'_i}{\vp_i}$
replaced by r.h.s.\ of~(\ref{2.12}) is valid for arbitrary~$\al_i$ we have{\samepage
 \begin{gather}
   \left(\ln R^2\frac{h}{H_i^2}\right)_{\!,i}=p_i, \qquad i=1,2,\dots,n   \label{2.13}
 \end{gather}
and from (\ref{2.13}) both (\ref{th1_1}) and (\ref{th1_2}) follow.}

Conditions (\ref{th1_1}) and (\ref{th1_2}) imply $R$-separability.

Indeed, we form the equations of~(\ref{def2}) with $p_i=\left(\ln R^2\frac{h}{H_i^2}\right)_{\!,i}$ and~$q_i$ given by~(\ref{th1_2}).
Then the implication~(\ref{def2}) in Def\/inition~\ref{definition2} is satisf\/ied.

{\bf B.} Indeed, the metric (\ref{metdiag}) satisf\/ies (\ref{th1_1}) if and only if there exist $2n$
functions $f_i(u^i)$
and $F_{(i)}(u^1,\dots,u^{i-1},u^{i+1},\dots,u^n)$ such that
 \begin{gather}
   \frac{h}{H_i^2}=\frac{1}{R^2}f_iF_{(i)}.  \label{hoverHi2}
 \end{gather}
Certainly, without loss of generality we can replace $F_{(i)}$ by
$\big(\prod\limits_{k\neq i} f_k\big)^{-1}G_{(i)}^2$, where
$G_{(i)}=G_{(i)}(u^1$, $\dots,u^{i-1},u^{i+1},\dots,u^n)$. Now (\ref{hoverHi2}) implies (\ref{2.16}) and
vice-versa.
 \end{proof}

 \begin{remark}\label{remark5}
Willard Miller Jr. derived (\ref{th1_1}) in \cite{WM2}. See (3.23) of \cite{WM2} and notice that his $R$ is
our $\ln R$.
 \end{remark}

Notice that  (\ref{2.16}) for $n=3$  gives the isothermic metric of  Def\/inition  4.

 \begin{definition}\label{definition5}
  The metric (\ref{2.16}) is called isothermic.
 \end{definition}

We introduce now an important sub-class of isothermic metrics. Given $n\choose 2$ functions
$G_{ij}=G_{ij}(u^i,u^j)$ $(i<j)$. We select $G_{(i)}$ as
follows
 \begin{gather*}
       G_{(i)}=\prod_{p\neq i \neq q} G_{pq}.   %\label{G(i)}
 \end{gather*}
Then (\ref{2.16}) assumes the form
 \begin{gather}
   ds^2=R^{4/(2-n)}\sum_{i=1}^n\frac{\prod\limits_{i<q}G_{iq}^2\prod\limits_{p<i}G_{pi}^2}{f_i^2}\big(du^i\big)^2.
     \label{binmet}
 \end{gather}

 \begin{definition}\label{definition6}
 The metric (\ref{binmet}) is called binary.
 \end{definition}

 \begin{example*}%\label{example1}
The $n$-elliptic coordinates on $\E^n$ \cite{CJ,KM2,MF,AU}. We choose $n$
real numbers $b_i$ such that  $b_1>b_2>\cdots >b_n$. The $n$-elliptic coordinates $\lambda=\left(\lambda^1,
\lambda^2,\dots,\lambda^n\right)$ satisfy inequalities
 \begin{gather*}
   \lambda^1>b_1>\lambda^2>\cdots >b_{n-1}>\lambda^n>b_n.
 \end{gather*}
The following formulae give rise to a dif\/feomorphism onto any of $2^n$  open $n$-hyper-octants of  $\E^n$
equipped with the standard Cartesian coordinates $x=(x^1,x^2,\dots,x^n)$
 \begin{gather*}
   \big(x^i\big)^2=\frac{\prod\limits_{j=1}^n\left(\la^j-b_i\right)}{\prod\limits_{j\neq i}\left(b_j-b_i\right)}, \qquad
i=1,2,\dots,n.
 \end{gather*}
The corresponding $n$-elliptic metric is
 \begin{gather}
   ds^2=\sum_{i=1}^n\frac{\prod\limits_{j\neq i}(\lambda^i-\lambda^j)}{4\prod\limits_{k=1}^n (\lambda^i-b_k)}(d\la^i)^2.
      \label{ellmet}
 \end{gather}

The $n$-elliptic metric is binary ($R=1$, $G_{ij}=\sqrt{\lambda^i-\lambda^j}$ and $f_i^2=4(-1)^{i-1}
\prod\limits_{k=1}^n(\lambda^i-b_k)$) and thus isothermic.
 \end{example*}

\subsection[$R$-equation]{$\boldsymbol{R}$-equation}\label{section2.2}

Having found the general formulae (\ref{2.16}) and (\ref{binmet}) for isothermic metrics
which~-- ex def\/initione~-- satisfy the 1st condition of $R$-separability, we are in a
position to claim that the various questions of $R$-separability amount to the 2nd
condition of $R$-separability (\ref{th1_2}) which we call $R$-equation.

 \begin{remark}\label{remark6}
(\ref{th1_2}) is not the Schr\"odinger equation since it is either a functional equation (when~$R$ is harmonic) or
$\Delta$  involves $R$.
 \end{remark}

 \begin{theorem}\label{theorem4}
 {\bf A.}  The metric \eqref{2.16} is $R$-separable in the Schr\"odinger equation if and only if
 \begin{gather}
   \sum_{i=1}^n G_{(i)}^2 f_i^2\! \left[\left(\frac{1}{R}\right)_{\!,ii}+
    \frac{f'_i}{f_i}\left(\frac{1}{R}\right)_{\!,i}+q_i\frac{1}{R}\right]\!=R^{(n+2)/(2-n)}
          G_{(1)}^{2/(n-2)}\cdots G_{(n)}^{2/(n-2)}\big(k^2-V\big).\!\!\!
             \label{eqn2}
 \end{gather}

{\bf B.} The binary metric \eqref{binmet} is $R$-separable in the  Schr\"odinger equation if and only if
 \begin{gather}
  \sum_{i=1}^n\frac{f_i^2}{\prod_{i<q}G_{iq}^2\prod_{p<i}G_{pi}^2}\left[\left(\frac{1}{R}\right)_{\!,ii}
+\frac{f'_i}{f_i}\left(\frac{1}{R}\right)_{\!,i}+q_i\frac{1}{R}\right]=R^{(n+2)/(2-n)}
\left(k^2-V\right). \label{th3}
\end{gather}
 \end{theorem}

 \begin{proof}
Indeed, both (\ref{eqn2}) and (\ref{th3}) are $R$-equations rewritten in terms of the corresponding metric.
 \end{proof}

 \begin{remark}\label{remark7}
 Notice that the linear operators acting on $R^{-1}$ in (\ref{eqn2}) and (\ref{th3}) also
def\/ine the separation equations (\ref{sepeqns}).
 \end{remark}

\subsection[B\^ocher-Ushveridze  identities]{B\^ocher--Ushveridze  identities}\label{section2.3}

Gaston Darboux was the f\/irst to discuss the so called triply conjugate coordinates in $\E^3$ \cite{D7}.
These constitute a projective generalization of orthogonal coordinates in $\E^3$. In this context he
introduced the following system of three equations for a single unknown $M(x_1, x_2, x_3)$
 \begin{gather}
 (x_1-x_2)M_{,12}-M_{,1}+M_{,2}=0, \nonumber\\
 (x_1-x_3)M_{,13}-M_{,1}+M_{,3}=0, \label{3dimEPD}\\
 (x_2-x_3)M_{,23}-M_{,2}+M_{,3}=0. \nonumber
 \end{gather}
and gave a general solution to it in the form
 \begin{gather}
    M=\frac{m_1(x_1)}{(x_1-x_2)(x_1-x_3)}+\frac{m_2(x_2)}{(x_2-x_1)(x_2-x_3)}+
      \frac{m_3(x_3)}{(x_3-x_1)(x_3-x_2)},   \label{3dimM}
 \end{gather}
where $m_i(x_i)$ are arbitrary functions
(see formulae (40), (41) and (42) in \cite{D7}).
A generalization of (\ref{3dimEPD})  and (\ref{3dimM}) is straightforward.

Consider in $\mathbb{R}^n$ the following system of $n \choose 2$ PDEs for a single unknown
$M(x_1,x_2,\dots,x_n)$
 \begin{gather}
     (x_i-x_j)M_{,ij}-M_{,i}+M_{,j}=0, \qquad i<j.   \label{EPD}
 \end{gather}
This is the overdetermined system of PDEs which is an example of the so called linear
Darboux--Manakov--Zakharov system~\cite{PV}. Fortunately (\ref{EPD}) is involutive (see
Proposition~1 in~\cite{PV}). Its general solution reads{\samepage
 \begin{gather*}
     M=\sum_{i=1}^n \frac{m_i(x_i)}{\prod\limits_{j\neq i}(x_i-x_j)},   %\label{M}
 \end{gather*}
where $m_i(x_i)$ are arbitrary functions.}

On the other hand each single equation of the system  (\ref{EPD}) is a particular case of the
Euler--Poisson--Darboux equation \cite[p.~54]{D9}  provided we ignore variables not
explicitly
involved in the equation. For simplicity  (\ref{EPD}) will be called the Euler--Poisson--Darboux
system.

 \begin{remark}\label{remark8}
Interestingly, in modern times the Euler--Poisson--Darboux system
and its various modif\/ications have been studied in the context of the so called integrable hydrodynamic
type systems \cite{ST1,ST2,MP}.
 \end{remark}

Notice particular solutions to (\ref{EPD}): $M = 0$, $M = 1$  and  $M =\sum\limits_{i=1}^n x_i$. The obvious
question arises as to what  functions $m_i(x_i)$ correspond to them.

   Maxime B\^ocher in his monograph on $R$-separability in the Laplace equation on $\E^n$  published without
proof a series of remarkable algebraic identities \cite[p.~250]{B}. These can be written in a~compact
form as follows
 \begin{gather}
   \sum^n_{i=1} \frac{x_i^{p-1}}{\prod\limits_{j\neq i}(x_i-x_j)}=\delta_{pn},
           \qquad  p=1,2,\dots,n.    \label{sum}
 \end{gather}

A.G.~Ushveridze generalized the identities (\ref{sum}) \cite{AU}.
We put $m=0,1,2,\dots$, $n=2,3,\dots$, $d=m+1-n$ and
 \begin{gather}
     f^{(n)}_d(x_1,x_2,\dots,x_n)=\sum^n_{i=1}\frac{x_i^m}{\prod\limits_{j\neq i}(x_i-x_j)}. \label{fnd1}
 \end{gather}
Then
 \begin{gather}
   f^{(n)}_d=\left\{\begin{array}{lll}
              0 & \textrm{for} & 0\leq m<n-1,\\
              1  & \textrm{for} & m=n-1,\\
             \textrm{homogeneous polynomial}  &&\\
             \textrm{of degree and homogeneity}=d &   \textrm{for} & m\geq n.
                    \end{array} \right. \label{fnd}
 \end{gather}
I.e. for $m \geq n$
 \begin{gather*}
     f^{(n)}_d=\sum_{1l_1+2l_2+\cdots+dl_d=d}f_{l_1l_2\dots l_d}\,\sigma^{l_1}_1
               \sigma^{l_2}_2\cdots\sigma^{l_d}_d,    %\label{fnd2}
 \end{gather*}
where $\sigma_i$ are elementary symmetric polynomials: $\sigma_1=\sum\limits_{i=1}^n x_i$,
$\sigma_2=\sum\limits_{i<j} x_i x_j$, $\dots$ and $f_{l_1l_2\dots l_d}$ are constants def\/ined
uniquely by r.h.s.\ of~(\ref{fnd1}). In particular
 \begin{gather}
   f^{(n)}_1=\sigma_1=\sum_{i=1}^n x_i,  \qquad
   f^{(n)}_2=\sigma_1^2-\sigma_2 ,\qquad f^{(n)}_3=\sigma_1^3-2\sigma_1\sigma_2+\sigma_3.\label{fn1}
 \end{gather}
The identities (\ref{fnd}) and in particular the identities (\ref{sum}) we call the B\^ocher--Ushveridze
identities. Certainly, both sides of any B\^ocher--Ushveridze identity is a particular solution to
the Euler--Poisson--Darboux  system. Notice also that functions $m_i(x_i)$ are not def\/ined by~$M$ uniquely.
As we shall see both the Euler--Poisson--Darboux  system and the B\^ocher--Ushveridze identities can be
applied in discussing $R$-equation.

\section{Examples}\label{section3}

In this section we discuss two standard results and two less standard results within the developed approach.
In all cases the approach of\/fers alternative and much simplif\/ied proofs or derivations.

\subsection{Robertson paper revisited}\label{section3.1}

Here we present the essence of Howard Percy Robertson fundamental paper~\cite{HR}. Our aim is to re-derive
the basic
formulae (A), (B), (C) and (9) of the paper using earlier stated results.

In (1) of \cite{HR} we put $k=1$ and replace $E$ by $k^2$. Notice that e.g.\ Robertson's $h_i$ is our
$H_i^{-2}$. The paper deals with the case $R=1$. $R$-equation (\ref{th1_2}) is now the
functional constraint which is bilinear in $H_i^{-2}$ and $q_i$
 \begin{gather}
     \sum_{i=1}^n \frac{1}{H_i^2}q_i=k^2-V .   \label{kV}
 \end{gather}
We decompose $q_i$ as follows
 \begin{gather}
   q_i\big(u^i\big)=k^2 q_{i1}\big(u^i\big)-v_i(u^i)+Q_i\big(u^i\big), \label{qi}
 \end{gather}
where $v_i$ are arbitrary. Inserting (\ref{qi}) into (\ref{kV}) gives
 \begin{gather}
    \sum_{i=1}^n \frac{1}{H_i^2}q_{i1}=1, \label{3eqn1}\\
    \sum_{i=1}^n \frac{1}{H_i^2}Q_i=0, \label{3eqn2}\\
    \sum_{i=1}^n \frac{1}{H_i^2}v_i=V.  \label{V}
 \end{gather}
Formally (\ref{3eqn2}) means that vector $(Q_i)$ belongs to $(n-1)$-dimensional orthogonal complement
of the vector $(H_i^{-2})$. Select a basis $(q_{ij})$ $(j=2,3,\dots,n)$ of the orthogonal complement
 \begin{gather}
   \sum_{i=1}^n \frac{1}{H_i^2}q_{ij}\big(u^i\big)=0,  \qquad j=2,3,\dots,n \label{3eqn3}
 \end{gather}
and decompose $(Q_i)$ in this basis as follows
 \begin{gather}
   Q_i\big(u^i\big)=\sum_{j=2}^n k_j q_{ij}\big(u^i\big),  \label{3eqn4}
 \end{gather}
where the coef\/f\/icients of the decomposition are arbitrary constants.

\begin{remark}\label{remark9}
(\ref{3eqn1}) and (\ref{3eqn3}) introduce (non-uniquely!) an $n\times n$ matrix $q=[q_{ij}(u^i)]$.
We assume that $q$ is non-singular everywhere. It is called the St\"ackel matrix. Notice that the co-factor
$Q_{ij}$ of $q_{ij}$ does not depend on~$u^i$.
\end{remark}

We collect (\ref{3eqn1}) and (\ref{3eqn3}) as
 \begin{gather}
   \sum_{i=1}^n\frac{1}{H_i^2}q_{ij}=\delta_{1j}. \label{3eqn5}
 \end{gather}
Inverting of (\ref{3eqn5}) yields
 \begin{gather}
   \frac{1}{H_i^2}=\left(q^{-1}\right)_{1i}=\frac{Q_{i1}}{\det q}.   \label{3eqn6}
 \end{gather}
It is clear that the metric
 \begin{gather}
  ds^2=\det q\sum_{i=1}^n\frac{(du^i)^2}{Q_{i1}}  \label{3eqn7}
 \end{gather}
satisf\/ies (\ref{kV}) or the 2-nd condition of $R$-separability (\ref{th1_2}).
Finally we demand the metric (\ref{3eqn7}) has to satisfy the f\/irst  condition of $R$-separability (\ref{th1_1})
 \begin{gather*}
   \left(\ln\frac{h}{H_i^2}\right)_{\!,ij}=\left(\ln\frac{h}{\det g}Q_{i1}\right)_{\!,ij}=
   \left(\ln\frac{h}{\det q}\right)_{\!,ij}=0, \qquad i\neq j,
 \end{gather*}
which implies
 \begin{gather}
     \frac{h}{\det q}=\prod_{i=1}^n f_i\big(u^i\big) \label{3eqn9}
 \end{gather}
and thus (\ref{hoverHi2}) is
 \begin{gather}
   \frac{h}{H_i^2}=f_i\big(u^i\big)Q_{i1}\prod_{j\neq i}f_j\big(u^j\big), \label{3eqn10}
 \end{gather}
which means that (\ref{3eqn10}) exactly conforms to (\ref{hoverHi2}).
As a result of (\ref{3eqn10}), (\ref{qi}) and (\ref{3eqn4}) the separation equations are
 \begin{gather}
   \vp''_i+\frac{f'_i}{f_i}\vp'_{i}+\left[k^2 q_{i1}+\sum_{j=2}^n k_j q_{ij}-v_i\right]\vp_i=0.
      \label{3eqn11}
 \end{gather}

To conclude we arrive at the following identif\/ications: (A), (B), (C) and (9)
of \cite{HR} are now (\ref{3eqn6}),  (\ref{V}), (\ref{3eqn9}) and (\ref{3eqn11}) respectively.

 \begin{definition}\label{definition7}
(\ref{3eqn7}) is called the St\"ackel metric and (\ref{3eqn9}) is called the Robertson condition.
 \end{definition}

\subsection[The $n$-elliptic metric]{The $\boldsymbol{n}$-elliptic metric}\label{section3.2}

It is well known that $n$-elliptic metric (\ref{ellmet}) is separable ($R = 1$)  in the Schr\"odinger equation
with an appropriately chosen potential function. An indirect proof consists in showing that~(\ref{ellmet})
is the St\"ackel metric (in this case the Robertson condition is satisf\/ied) and Eisenhart stated it without
proof in \cite[p.~302]{LE2}.

 \begin{theorem}\label{theorem5}
The $n$-elliptic metric \eqref{ellmet} is separable in the Schr\"odinger equation with a potential
function
 \begin{gather*}
  V(\lambda)= \sum_{i=1}^n\frac{v_i(\lambda^i)}{\prod\limits_{j\neq i}(\lambda^i-\lambda^j)}, %\label{th3_V}
 \end{gather*}
where  $v_i(\la^i)$ are arbitrary functions, i.e.\ $V(\la)$ is an arbitrary solution to the
Euler--Poisson--Darboux system  \eqref{EPD}. The corresponding separation equations are
 \begin{gather*}
  \vp''_i+\frac{1}{2}\frac{a'_i}{a_i}\vp'_i+\frac{1}{a_i}\left[\sum_{m=0}^{n-2}
     k_m(\lambda^i)^m+k^2(\lambda^i)^{n-1}-v_i(\lambda^i)\right]\vp_i=0, \qquad i=1,2,\dots,n,
 \end{gather*}
where  $a_i  = 4 \prod\limits_{k=1}^n(\la^i-b_k)$ and  $k_m$ are arbitrary constants $(m = 0, 1, \dots, n-2)$.
 \end{theorem}
 \begin{proof}
Indeed, from example~(Subsection~\ref{section2.1})  we know that the metric (\ref{ellmet}) is isothermic. Again $R$-equation is
reducible to the functional constraint
 \begin{gather*}
 \sum_{i=1}^n \frac{a_i(\lambda^i)q_i(\lambda^i)}{\prod\limits_{j\neq i}(\lambda^i-\lambda^j)}=
          k^2-V(\lambda).  %\label{th3_Req}
 \end{gather*}
We put
 \begin{gather*}
     a_i(\lambda^i)q_i(\lambda^i)=\sum_{m=0}^{n-2}k_m(\lambda^i)^m+
        k^2(\lambda^i)^{n-1}-v_i(\lambda^i), \qquad i=1,2,\dots,n,
 \end{gather*}
where $k_m={\rm const}$ and $v_i(\la^i)$ are arbitrary functions.
Now  the B\^ocher--Ushveridze identity~(\ref{sum}) implies the statement.
 \end{proof}

\subsection[Remarkable example of Kalnins-Miller revisited]{Remarkable example of Kalnins--Miller revisited}\label{section3.3}

Our setting can be easily
extended to the pseudo-Riemannian case. Consider the following metric
 \begin{gather}
   d\sigma^2= \left(\lambda^1-\lambda^2\right)\left(\lambda^1-\lambda^3\right)
    (d\lambda^1)^2+\left(\lambda^2-\lambda^1\right)\left(\lambda^2-\lambda^3\right)
    (d\lambda^2)^2\nonumber\\
\phantom{d\sigma^2=}{} +\left(\lambda^3-\lambda^1\right)
      \left(\lambda^3-\lambda^2\right)(d\lambda^3)^2,
 \label{dsigma2}
 \end{gather}
where $\la^1>\la^2>\la^3>0$.
It is 3-dimensional Minkowski metric. Indeed,
on replacing $\lambda^i$ by~$t$,~$x$ and $y$
 \begin{gather*}
   t=\frac{1}{9}\left(\la^1+\la^2+\la^3\right)-\frac{9}{16}\left(\la^1+\la^2-\la^3\right)
     \left(\la^1-\la^2+\la^3\right)\left(\la^1-\la^2-\la^3\right),\\
   x=\frac{1}{9}\left(\la^1+\la^2+\la^3\right)+\frac{9}{16}\left(\la^1+\la^2-\la^3\right)
     \left(\la^1-\la^2+\la^3\right)\left(\la^1-\la^2-\la^3\right),\\
   y=\frac{1}{4}\left(\la^1+\la^2-\la^3\right)^2-\la^1\la^2,
 \end{gather*}
we arrive at
 \begin{gather*}
     d\sigma^2=-d t^2+d x^2 +d y^2.
 \end{gather*}
Certainly, any metric conformally equivalent to (\ref{dsigma2}) is an isothermic
metric and thus satisf\/ies the 1st condition of $R$-separability (\ref{th1_1}).
Kalnins and Miller proved that the metric
 \begin{gather}
    d s^2=\left(\la^1+\la^2+\la^3\right)d\sigma^2  \label{KMmet}
 \end{gather}
is $R$-separable in the Helmholtz equation (\ref{Helm})  \cite[p.~472]{KM1}. We re-derive this
remarkable result within our approach.

First of all it is easy to predict $R$-factor (see (\ref{binmet})) and the form of the separation
equations $(f_1^2=f_3^2=1$, $f_2^2=-1)$
 \begin{gather}
     R(\la)=\left(\la^1+\la^2+\la^3\right)^{-1/4}, \label{R}\\
     \vp''_i+q_i\vp_i = 0, \qquad i=1,2,3. \nonumber
 \end{gather}
The point is that (\ref{R}) is harmonic with respect of (\ref{KMmet}). Again
$R$-equation is reducible to the functional constraint
 \begin{gather*}
  \left(\la^1+\la^2+\la^3\right)k^2=\frac{q_1(\la^1)}
                       {\left(\lambda^1-\lambda^2\right)\left(\lambda^1-\lambda^3\right)}+
    \frac{q_2(\la^2)}{\left(\lambda^2-\lambda^1\right)\left(\lambda^2-\lambda^3\right)}+
    \frac{q_3(\la^3)}{\left(\lambda^3-\lambda^1\right)\left(\lambda^3-\lambda^2\right)}.
 \end{gather*}
Then from the B\^ocher--Ushveridze identities (\ref{sum}) and (\ref{fn1}) we have immediately
 \begin{gather*}
   q_i(\la^i)=k^2(\la^i)^3+k_1(\la^i)+k_0,
 \end{gather*}
where $k_0$, $k_1$ are arbitrary constants.

\subsection[Fixed energy $R$-separation revisited]{Fixed energy $\boldsymbol{R}$-separation revisited}\label{section3.4}

In order to treat $R$-separability in the Schr\"odinger equation (\ref{Schrk0}) a pretty complicated
formalism was proposed  in~\cite{CR}. Presumably some part of the
formalism of~\cite{CR} can be simplif\/ied according to the following result.

 \begin{proposition}\label{proposition1}
Any  isothermic metric which is $R$-separable in $n$-dimensional Laplace equation is $R$-separable in $n$-dimensional
Schr\"odinger equation with $k = 0$ for an appropriately chosen potential function.
 \end{proposition}

 \begin{proof}
Consider the isothermic metric given by (\ref{2.16}).
$R$-separability of (\ref{2.16}) in $n$-dimensional Laplace equation implies that $R$-equation simplif\/ies to
 \begin{gather}
   R^{-1}\Delta R-\sum_{i=1}^n\frac{1}{H_i^2}q_i=0. \label{3.4_Req}
 \end{gather}
We put
 \begin{gather}
   V(u)=\sum_{i=1}^n\frac{1}{H_i^2}v_i\big(u^i\big),  \label{3.4_V}
 \end{gather}
where $v_i(u^i)$ are arbitrary functions and def\/ine $\bar{q}_i=q_i-v_i$. Then  (\ref{3.4_Req}) can be
rewritten as
 \begin{gather*}
   R^{-1}\Delta R-\sum_{i=1}^n\frac{1}{H_i^2}v_i-\sum_{i=1}^n\frac{1}{H_i^2}\bar{q}_i=0,
 \end{gather*}	
which is $R$-equation  for  $n$-dimensional Schr\"odinger equation  (\ref{Schrk0}) with the potential func\-tion~(\ref{3.4_V}).
 \end{proof}

\section[$R$-separability in  3-dimensional case]{$\boldsymbol{R}$-separability in  3-dimensional case}\label{section4}

\subsection[Procedure to detect $R$-separable metrics]{Procedure to detect $\boldsymbol{R}$-separable metrics}\label{section4.1}

Here we describe a simple procedure to identify a given 3-dimensional diagonal metric as $R$-separable
in 3-dimensional Laplace equation.

 \begin{proposition}\label{proposition2}
In  the $3$-dimensional case any isothermic metric is binary.
 \end{proposition}

 \begin{proof}
Indeed, we put $n=3$ in (\ref{2.16}) and hence we deduce the following expressions for Lam\'e coef\/f\/icients  $H_i$
 \begin{gather*}
     H_i=\frac{1}{R^2}G_{(1)}G_{(2)}G_{(3)}G_{(i)}^{-1}\frac{1}{f_i}, \qquad i=1,2,3,
 \end{gather*}
or more explicitly
 \begin{gather}
  H_1=\frac{G_{(2)}G_{(3)}}{R^2f_1}=\frac{G_{12}G_{13}}{R^2f_1}, \qquad   H_2=\frac{G_{(1)}G_{(3)}}{R^2f_2}=
  \frac{G_{12}G_{23}}{R^2f_2}, \nonumber\\
  H_3=\frac{G_{(1)}G_{(2)}}{R^2f_3}=\frac{G_{13}G_{23}}{R^2f_3}.
       \label{Hi}
 \end{gather}
Now see (\ref{binmet}).
 \end{proof}

To simplify notation we rewrite  (\ref{Hi}) as
  \begin{gather*}
  H_1=\frac{G_{2}G_{3}}{Mf_1}, \qquad   H_2=\frac{G_{1}G_{3}}{Mf_2}, \qquad  H_3=\frac{G_{1}G_{2}}{Mf_3}.
       %\label{Hi1}
 \end{gather*}
In other words  $G_i$ does not depend  on $u^i$, $f_i$ depends  on $u^i$ and  $R=\sqrt{M}$. Finally we arrive at
the following general form of the isothermic metric in 3-dimensional case
 \begin{gather}
   ds^2=\frac{1}{M^2}\left[\frac{G_2^2G_3^2}{f_1^2}\big(du^1\big)^2+\frac{G_1^2G_3^2}{f_2^2}\big(du^2\big)^2+
            \frac{G_1^2G_2^2}{f_3^2}\big(du^3\big)^2\right].    \label{3dim_met}
 \end{gather}

The procedure in question consists of  three steps. Suppose we are given any 3-dimensional diagonal metric
 \begin{gather}
   ds^2=H_1^2(u)\big(du^1\big)^2+H_2^2(u)\big(du^2\big)^2+H_3^2(u)\big(du^3\big)^2.   \label{3dimdiagmet}
 \end{gather}
In the f\/irst step we attempt to identify (\ref{3dimdiagmet}) as an isothermic metric (\ref{3dim_met}).
Suppose it is the case.
This step provides us with (predicts) possible forms of  $R$-factor  and coef\/f\/icients $p_i$ in the separation equations.

In the second step we form the $R$-equation (\ref{th1_2}) for 3-dimensional Laplace equation either in terms of (\ref{3dimdiagmet})
as
 \begin{gather}
   R^{-1}\Delta R-\sum_{i=1}^3\frac{1}{H_i^2}q_i=0,  \label{4.6}
 \end{gather}
or in terms of (\ref{3dim_met}) as
 \begin{gather}
   \Delta R-\left(\frac{s_1}{G_2^2G_3^2}+\frac{s_2}{G_1^2G_3^2}+\frac{s_3}{G_1^2G_2^2}\right)R^5=0, \label{4.7}
 \end{gather}
where $s_i = f_i^2q_i$ or as
  \begin{gather}
         \sum_{i=1}^3 G_{i}^2f_i^2\left[\p^2_iR^{-1}+\frac{f'_i}{f_i}\p_i R^{-1}+q_i R^{-1} \right]=0 .
         \label{4.8}
  \end{gather}

In the third step we treat  (\ref{4.6}) and  (\ref{4.8}) as equations for unknowns $q_i(u^i)$ and (\ref{4.7}) as
equation for unknowns $s_i(u^i)$. Any solution to (\ref{4.6}), (\ref{4.7}) or (\ref{4.8}) provides us with a
coef\/f\/icient $q_i$ in the separation equations. If the third step is successful, then the starting metric
(\ref{3dimdiagmet}) is $R$-separable in 3-dimensional Laplace equation and the separation equations are constructed explicitly.

If $R$ is harmonic with respect of (\ref{3dim_met}) (this case includes separability), then e.g.~(\ref{4.6})
becomes a linear in $q_i$ constraint
 \begin{gather}
   \sum_{i=1}^3\frac{1}{H_i^2}q_i=0  \label{4.9}
 \end{gather}
and the corresponding solution space is at most 2-dimensional.

If $R$ is not harmonic with respect of (\ref{3dim_met}) and if e.g.~(\ref{4.6}) admits a special solution
$q_{i0}$, then a general solution to  (\ref{4.6}) is
 \begin{gather*}
  q_i=q_{i0}+q_{i1},
 \end{gather*}
where $q_{i1}$ is a solution to  (\ref{4.9}).

\subsection{Examples}\label{section4.2}

Here we present f\/ive examples proving ef\/f\/iciency of our procedure.

\subsubsection{Spherical metric}\label{section4.2.1}
The spherical metric
 \begin{gather*}
   ds^2=dr^2+r^2d\theta^2+r^2\sin^2\theta d\phi^2
 \end{gather*}
is isothermic. It is easily seen that
 \begin{gather*}
   R=1,\qquad G_1=\sin\theta,\qquad G_2=r,\qquad G_3=r,\qquad f_1=r^2,\qquad f_2=\sin\theta,\qquad f_3=1
 \end{gather*}
in this case.
Equation~(\ref{4.9}) reads
 \begin{gather*}
   q_1+\frac{1}{r^2}q_2+\frac{1}{r^2\sin^2\theta}q_3=0
 \end{gather*}
and can be easily solved
 \begin{gather*}
   q_1=-\frac{\alpha}{r^2},\qquad q_2=\alpha-\frac{\beta}{\sin^2\theta},\qquad q_3=\beta, \qquad \alpha,\beta={\rm const}.
 \end{gather*}
The resulting separation equations read
 \begin{gather*}
  \vp''_1+\frac{2}{r}\vp'_1-\frac{\alpha}{r^2}\vp_1=0, \nonumber\\
  \vp''_2+\cot\theta\ \vp'_2+\left(\alpha-\frac{\beta}{\sin^2\theta}\right)\vp_2=0,\\
  \vp''_3+\beta\vp_3=0. \nonumber
 \end{gather*}

\subsubsection{Toroidal metric I}\label{section4.2.2}

The so called  toroidal metric
 \begin{gather}
  ds^2=\left(\cosh\eta-\cos\theta\right)^{-2}\left(d\eta^2+d\theta^2+\sinh^2\eta\, d\phi^2\right), \label{tormet}
 \end{gather}
discussed in e.g.~\cite{MS,CR}, is isothermic and
 \begin{gather*}
   R=\sqrt{\cosh\eta-\cos\theta},\qquad G_1=G_3=1,\qquad G_2=\sinh\eta,\\ f_1=\sinh\eta,\qquad f_2=f_3=1.
 \end{gather*}
We easily verify the equality
 \begin{gather}
  \Delta R-\frac{1}{4}R^5=0.   \label{torReq}
 \end{gather}
Hence  equation (\ref{4.7}) is satisf\/ied if and only if
 \begin{gather}
   \frac{1}{\sinh^2\eta}s_1+s_2+\frac{1}{\sinh^2\eta}s_3=\frac{1}{4}.  \label{4.2.2eqn}
 \end{gather}
A general solution to (\ref{4.2.2eqn}) is
 \begin{gather*}
   s_1=f_1^2q_1=\left(\frac{1}{4}-\alpha_1\right)\sinh^2\eta-\alpha_2,\\ s_2=f_2^2q_2=\alpha_1,\qquad
   s_3=f_3^2q_3=\alpha_2, \qquad  \alpha_1,\alpha_2={\rm const}.
 \end{gather*}
The resulting separation equations read
 \begin{gather*}
  \vp''_1+\coth\eta\ \vp'_1+\left(\frac{1}{4}-\alpha_1-\frac{1}{\sinh^2\eta}\alpha_2\right)\vp_1=0, \nonumber\\
  \vp''_2+\alpha_1\vp_2=0,\\
  \vp''_3+\alpha_2\vp_3=0. \nonumber
 \end{gather*}

\subsubsection{Toroidal metric II}\label{section4.2.3}

Interestingly, the metric (\ref{tormet}) can be identif\/ied as isothermic in two ways.  It was shown implicitly
in~\cite{CR}. Indeed, we rewrite (\ref{tormet}) as follows
 \begin{gather}
  ds^2=\frac{\sinh^2\eta}{(\cosh\eta-\cos\theta)^2}\left(\frac{d\eta^2+d\theta^2}{\sinh^2\eta}+d\phi^2\right).
                \label{tormet1}
 \end{gather}
Metric (\ref{tormet1}) suggests the following identif\/ications
 \begin{gather*}
   R=\sqrt{\coth\eta-\frac{\cos\theta}{\sinh\eta}},\qquad G_1=G_2=1,\qquad G_3=\frac{1}{\sinh\eta},\qquad
    f_1=f_2=f_3=1.
 \end{gather*}
Again (\ref{torReq}) holds. Hence equation (\ref{4.7}) is satisf\/ied if and only if
 \begin{gather}
  s_1\sinh^2\eta+s_2\sinh^2\eta+s_3=\frac{1}{4}.  \label{4.2.3eqn}
 \end{gather}
A general solution to (\ref{4.2.3eqn}) is
 \begin{gather*}
   s_1=\left(\frac{1}{4}-\alpha_2\right)\frac{1}{\sinh^2\eta}-\alpha_1,\qquad s_2=\alpha_1,\qquad
   s_3=\alpha_2, \qquad  \alpha_1,\alpha_2={\rm const}.
 \end{gather*}
The resulting separation equations read
 \begin{gather*}
  \vp''_1+\left[\left(\frac{1}{4}-\alpha_2\right)\frac{1}{\sinh^2\eta}-\al_1\right]\vp_1=0, \nonumber\\
  \vp''_2+\alpha_1\vp_2=0,\\
  \vp''_3+\alpha_2\vp_3=0. \nonumber
 \end{gather*}

\subsubsection{Cyclidic metric}\label{section4.2.4}

Consider the following  metric
 \begin{gather}
  ds^2=\left(1+p\sqrt{\la^1\la^2\la^3}\right)^{-2}\left[\frac{(\la^1-\la^2)(\la^1-\la^3)(d\la^1)^2}
       {\vp(\la^1)}+\frac{(\la^2-\la^1)(\la^2-\la^3)(d\la^2)^2}{\vp(\la^2)}\right.\nonumber\\
 \left.\phantom{ds^2=}{}
 +\frac{(\la^3-\la^1)(\la^3-\la^2)(d\la^3)^2}{\vp(\la^3)}\right],
 \label{cycmet}
 \end{gather}
where $\vp(x)=(x-a)(x-b)(x-c)(x-d)$ and $p$, $a$, $b$, $c$, $d$ are constants. In general it is not f\/lat.

 \begin{proposition}\label{proposition3}
The off-diagonal components of Ricci tensor of \eqref{cycmet} vanish, i.e.\ part of Lam\'e equations \eqref{Lameoffdiag}
is satisfied. The diagonal components of Ricci tensor of \eqref{cycmet} vanish, i.e.\ the other part of Lam\'e equations
\eqref{Lamediag} is satisfied, if and only if
 \begin{gather*}
  pabcd=0 \qquad \text{and} \qquad p^2(abc+abd+acd+bcd)=1.
 \end{gather*}
 \end{proposition}

We select  $d = 0$ and $p = 1/\sqrt{abc}$. Hence the metric
 \begin{gather}
  ds^2=\left(1+\sqrt{\frac{\la^1\la^2\la^3}{abc}}\right)^{-2}\left[\frac{(\la^1-\la^2)(\la^1-\la^3)(d\la^1)^2}
       {\vp(\la^1)}+\frac{(\la^2-\la^1)(\la^2-\la^3)(d\la^2)^2}{\vp(\la^2)}\right.\nonumber\\
      \left.\phantom{ds^2=}{} +\frac{(\la^3-\la^1)(\la^3-\la^2)(d\la^3)^2}{\vp(\la^3)}\right]
\label{cycmet1}
 \end{gather}
with  $\vp(x)  =  x(x-a)(x-b)(x-c)$ is f\/lat. It is isothermic and
 \begin{gather*}
   R^2=\left(1+\sqrt{\frac{\la^1\la^2\la^3}{abc}}\right),\qquad G_1^2=\la^2-\la^3,\qquad G_2^2=\la^1-\la^3,\qquad
   G_3^2=\la^1-\la^2, \\
   f_1^2=\vp(\la^1),\qquad f_2^2=-\vp(\la^2), \qquad f_3^2=\vp(\la^3).
 \end{gather*}
We readily check the equality
 \begin{gather*}
  \Delta R-\frac{3}{16}R^5=0.
 \end{gather*}
Hence equation (\ref{4.7}) is satisf\/ied if and only if
 \begin{gather}
 \frac{\vp(\la^1)q_1}{(\la^1-\la^2)(\la^1-\la^3)}+\frac{\vp(\la^2)q_2}{(\la^2-\la^1)(\la^2-\la^3)}+
    \frac{\vp(\la^3)q_3}{(\la^3-\la^1)(\la^3-\la^2)}=\frac{3}{16}.   \label{4.2.4eqn}
 \end{gather}
From B\^ocher--Ushveridze identities (\ref{sum})  $(n = 3)$  we deduce immediately a general solution to~(\ref{4.2.4eqn})	
 \begin{gather*}
  q_i=\frac{1}{\vp(\la^i)}\left(\al_1+\al_2\la^i+\frac{3}{16}\big(\la^i\big)^2\right), \qquad i=1,2,3,
 \end{gather*}	
where $\al_1,\al_2={\rm const}$.
The resulting separation equations read
 \begin{gather*}
 \vp''_i+\frac{1}{2}\frac{\vp'(\la^i)}{\vp(\la^i)}\vp'_i + \frac{1}{\vp(\la^i)}\left[\al_1+\al_2\la^i+
     \frac{3}{16}\big(\la^i\big)^2\right]\varphi_i=0.
 \end{gather*}

 \begin{definition}\label{definition8}
A diagonal 3-dimensional f\/lat metric all whose parametric surfaces are cyclides (Dupin cyclides) is called
cyclidic (Dupin-cyclidic).
 \end{definition}

General cyclides are discussed in \cite{NT}. For Dupin cyclides see Section~\ref{section4.3} of the paper.

Metric (\ref{cycmet1}) is cyclidic but not Dupin-cyclidic.

\subsubsection{Dupin-cyclidic metric}\label{section4.2.5}

The metric
 \begin{gather}
 ds^2=\frac{b^2(w-a\cosh v)^2}{(a\cosh v-c\cos u)^2}du^2+
            \frac{b^2(w-c\cos u)^2}{(a\cosh v-c\cos u)^2}dv^2+dw^2  \label{dupcycmet}
 \end{gather}
is Dupin-cyclidic \cite{PS}. It is $R$-separable in the Helmholtz equation (\ref{Helm}) on~$\E^3$
(see Theorem~1 in~\cite{PS}). Here we give an alternative and remarkably simple proof of this result.
Metric~(\ref{dupcycmet}) is isothermic and
 \begin{gather*}
   R=(a \cosh v-w)^{-1/2}(w-c \cos u)^{-1/2},\qquad  G_1 =(a \cosh v -w)^{-1},\\ G_2=(w- c \cos u)^{-1},\qquad
   G_3 = (a \cosh v-c \cos u)^{-1},\qquad f_1 = f_2 = b^{-1},\qquad f_3 = 1.
 \end{gather*}
It is easily to verify the equality
 \begin{gather}
  R^{-1} \Delta R- \frac{1}{4}\left(H_1^{-2} - H_2^{-2}\right) = 0,    \label{dupcycReq}
 \end{gather}
which is equation (\ref{4.6}) in this case. Taking into account $H_3 = 1$ we rewrite~(\ref{dupcycReq})
 \begin{gather}
   R^{-1} \Delta R+k^2- \frac{1}{4}\left(H_1^{-2} - H_2^{-2}\right)-k^2 H_3^{-2} = 0.    \label{dupcycReq1}
 \end{gather}
Certainly, (\ref{dupcycReq1}) is $R$-equation (\ref{th1_2}) for 3-dimensional Helmholtz equation (\ref{Helm}). The corresponding separation
equations are
 \begin{gather*}
  \vp''_1+\frac{1}{4}\vp_1 = 0,\\
  \vp''_2-\frac{1}{4}\vp_2 = 0,\\
  \vp''_3+k^2\vp_3 = 0.
 \end{gather*}

\subsection{Dupin-cyclidic metrics}\label{section4.3}

Gaston Darboux found a broad class of Dupin-cyclidic metrics which are $R$-separable in 3-dimensional Laplace equation
\cite[Section~162, p.~286]{D8}. Here we give an alternative and simplif\/ied proof of this remarkable
result. The metric (\ref{dupcycmet}) belongs to this class.

There are many def\/initions (not necessarily equivalent) of Dupin cyclides (see \cite[p.~148]{TC}).
We select the following one.

 \begin{definition}\label{definition9}
A Dupin cyclide is a  regular parametric surface in $\E^3$ whose
both principal curvatures are constant along their curvature lines.
 \end{definition}

Let us recall the celebrated theorem of Dupin (see \cite[p.~609]{GAS}).
 \begin{theorem}\label{theorem6}  Let  $u = (u^1, u^2, u^3)$  be orthogonal coordinates in $\E^3$. Two arbitrary parametric
surfaces $u^i = {\rm const}$  and  $u^j = {\rm const}$  $(i\neq j)$ intersect in a curvature line of each.
 \end{theorem}

 \begin{proposition}\label{proposition4}
The metric \eqref{3dimdiagmet}
is Dupin-cyclidic $($see Definition~{\rm 8)} if and only if  it is flat and its Lam\'e coefficients satisfy the
following six PDEs
 \begin{gather}
  \frac{\p}{\p u^j}H_i^{-1}\frac{\p}{\p u^i}\ln H_j =0 \qquad i,j=1,2,3, \quad i\neq j.  \label{prop4_eqns}
 \end{gather}
 \end{proposition}

 \begin{proof}
Indeed, $k_{ij}= -H_i^{-1}\frac{\p}{\p u^i}\ln H_j $ is a principal curvature on a parametric surface
$u^i = {\rm const}$ in the direction of a curvature line $u^j$-variable  \cite[p.~608]{GAS}.
 \end{proof}

A natural question arises as to when the isothermic metric (\ref{3dim_met}) satisf\/ies (\ref{prop4_eqns})?
With no dif\/f\/iculty we prove the following result.

 \begin{proposition}\label{proposition5}
The isothermic metric \eqref{3dim_met} satisfies \eqref{prop4_eqns} if and only if
  \begin{gather}
 \left(\frac{M}{G_1}\right)_{\!,23}=\left(\frac{M}{G_2}\right)_{\!,13}=
    \left(\frac{M}{G_3}\right)_{\!,12}=0.  \label{lem4}
 \end{gather}
  \end{proposition}

Certainly, one solution to (\ref{lem4}) is provided by the metric (\ref{dupcycmet}). On performing
re-scaling in (\ref{dupcycmet})
 \begin{gather*}
     u^1=c\cos u, \qquad u^2=a\cosh v,\qquad u^3=w
 \end{gather*}
we  arrive at
 \begin{gather}
   ds^2=H_1^2(u)\big(du^1\big)^2+H_2^2(u)\big(du^2\big)^2+H_3^2(u)\big(du^3\big)^2 \label{4.3met}
 \end{gather}
with
 \begin{gather}
   H_1=\big(u^3-u^1\big)\big(u^3-u^2\big)\big(u^1-u^2\big)^{-1}\big(u^1-u^3\big)^{-1}\left[-\frac{(u^1)^2}{b^2}+
       \frac{c^2}{b^2}\right]^{-1/2},  \label{4.3H1}\\
   H_2=\big(u^3-u^1\big)\big(u^3-u^2\big)\big(u^2-u^1\big)^{-1}\big(u^2-u^3\big)^{-1}\left[\frac{(u^2)^2}{b^2}-
       \frac{a^2}{b^2}\right]^{-1/2}, \\
   H_3=\big(u^3-u^1\big)\big(u^3-u^2\big)\big(u^3-u^1\big)^{-1}\big(u^3-u^2\big)^{-1}. \label{4.3H3}
 \end{gather}
Obviously, (\ref{4.3met}) is isothermic and from (\ref{4.3H1})--(\ref{4.3H3})  we have
 \begin{gather}
   G_1=\big(u^2-u^3\big)^{-1}, \qquad  G_2=\big(u^1-u^3\big)^{-1}, \qquad  G_3=\big(u^1-u^2\big)^{-1}. \label{Gs}
 \end{gather}
Let us insert (\ref{Gs}) into (\ref{lem4}). The resulting system of equations
 \begin{gather*}
  \big(u^i-u^j\big)M_{,ij}-M_{,i}+M_{,j}=0  \qquad i,j=1,2,3,\quad i<j
 \end{gather*}
is exactly the Euler--Poisson--Darboux system (\ref{EPD}) for $n = 3$. Hence $M$  is of the form
 \begin{gather}
 M=\frac{b_1(u^1)}{(u^1-u^2)(u^1-u^3)}+\frac{b_2(u^2)}{(u^2-u^1)(u^2-u^3)}+
     \frac{b_3(u^3)}{(u^3-u^1)(u^3-u^2)}, \label{5.23}
 \end{gather}
where $b_i(u^i)$ are arbitrary functions of a single variable. Thus we have the following result.

 \begin{lemma}\label{lemma1}
Any metric \eqref{4.3met}
with
 \begin{gather}
    H_i=\frac{1}{M}\big(u^i-u^j\big)^{-1}\big(u^i-u^k\big)^{-1}a_i^{-1/2}, \qquad i,j,k  \ \ \textrm{are different},
      \label{5.25}
 \end{gather}
where $M$  is given by  \eqref{5.23} while  $b_i(u^i)$ and  $a_i(u^i)$  are arbitrary
functions of a single variable is isothermic and satisfies \eqref{prop4_eqns}.
 \end{lemma}

 \begin{theorem}\label{theorem7}
Suppose a $3$-dimensional Riemann space  $\R^3$  admits   the metric  described in Lem\-ma~{\rm \ref{lemma1}}.
Then
 \begin{itemize}\itemsep=0pt
\item[$1)$] the off-diagonal  components of the Ricci tensor vanish,
\item[$2)$] $\R^3$ is  conformally flat  if and only if
  \begin{gather}
    a_i\big(u^i\big)=m_i\big(u^i\big)^2+2n_iu^i+p_i, \qquad  m_i,n_i,p_i={\rm const},  \label{116a}
  \end{gather}
where
 \begin{gather}
    \sum_{i=1}^3 m_i= \sum_{i=1}^3 n_i= \sum_{i=1}^3 p_i=0,   \label{116b}
 \end{gather}
\item[$3)$] $\R^3$ is flat  if and only if it is conformally flat and the following identities hold
 \begin{gather}
   \big(n_i^2-m_ip_i\big)b_i^2+2[(\beta_im_i-\alpha_i n_i)u^i+\beta_in_i-\alpha_i p_i]b_i+
   \big(\alpha_i u^i+\beta_i\big)^2+\gamma_i a_i =0, \label{117}
 \end{gather}
   where the constants $\alpha_i$, $\beta_i$ and $\gamma_i$ satisfy identities
 \begin{gather}
    \sum_{i=1}^3 \alpha_i= \sum_{i=1}^3 \beta_i= \sum_{i=1}^3 \gamma_i=0.  \label{118}
 \end{gather}
\end{itemize}
 \end{theorem}

 \begin{proof}
1) Suppose  we are given the metric (\ref{4.3met}) whose Lam\'e coef\/f\/icients are (\ref{5.25}) and
$M$ is an arbitrary function. Then of\/f-diagonal components of its Ricci tensor are
 \begin{gather*}
   R_{ij}=\frac{1}{M}\big(u^i-u^j\big)^{-1}\left[\big(u^i-u^j\big)M_{,ij}+M_{,j}-M_{,i}\right], \qquad i<j.
 \end{gather*}

2) The Cotton--York of  the metric vanishes if and only if  (\ref{116a}) and (\ref{116b}) hold.

3)
Suppose (\ref{116a}), (\ref{116b}) are valid, then the diagonal components of the Ricci
tensor vanish if and only if (\ref{117}) and (\ref{118}) hold.
 \end{proof}

 \begin{remark}\label{remark10}
The result  3) of Theorem~\ref{theorem7} is essentially due to Gaston Darboux. See the
formulae  in \cite[p.~335]{D7}.
 \end{remark}

 \begin{lemma}\label{lemma2}
Suppose $\E^3$ is equipped with the metric described in $3)$ of  Theorem~{\rm \ref{theorem7}}. Then
$R =\sqrt{M}$ satisfies the equation
 \begin{gather}
   \big(u^2-u^3\big)^{-2}\sqrt{a_1}\left[\p_1\sqrt{a_1}\p_1 R^{-1}-\frac{m_1}{4\sqrt{a_1}}
    R^{-1}\right]\nonumber\\
\qquad{}  +\big(u^1-u^3\big)^{-2}\sqrt{a_2}\left[\p_2\sqrt{a_2}\p_2 R^{-1}-\frac{m_2}{4\sqrt{a_2}}
    R^{-1}\right] \nonumber\\
 \qquad{}  +\big(u^1-u^2\big)^{-2}\sqrt{a_3}\left[\p_3\sqrt{a_3}\p_3 R^{-1}-\frac{m_3}{4\sqrt{a_3}}
    R^{-1}\right]=0. \label{5.31}
 \end{gather}
 \end{lemma}

We identify  (\ref{5.31}) as $R$-equation  (\ref{4.8}) for the Laplace equation on  $\E^3$.

 \begin{theorem}\label{theorem8}
Any Dupin-cyclidic metric  described in $3)$ of  Theorem~{\rm \ref{theorem7}} is $R$-separable in the
Laplace equation on $\E^3$. $R =\sqrt{M}$   and   $M$  is given by  \eqref{5.23}.
The separation equations are
 \begin{gather*}
   \vp''_i+\frac{1}{2}\frac{a'_i}{a_i}\vp'_i-\frac{m_i}{4a_i}\vp_i=0, \qquad i=1,2,3.
 \end{gather*}
 \end{theorem}

 \begin{remark}\label{remark11}
Theorem~\ref{theorem8} is due to Gaston Darboux as well. See his remarkable result  pointed
out in  \cite[Section~162, p.~286]{D8}.
 \end{remark}

\subsection*{Acknowledgments}

Our thanks are due to reviewers for critical remarks and notably to the editors for valuable comments which inspired us to deeply revise our preprint.

\pdfbookmark[1]{References}{ref}
\LastPageEnding

\end{document}